\documentclass[agupp]{aguplus}
\usepackage{graphics}

\sectionnumbers

\doublecaption{35pc}

\authoraddr{H. Z. Baumert, IAMARIS, Bei den M{\"u}hren 69A,
D-20457 Hamburg, Germany (baumert@iamaris.org)}

\usepackage{graphicx}
\begin{document}
\bibliographystyle{ametsoc}

\def\lesssim{\mathrel{\hbox{\rlap{\hbox{\lower0.45em\hbox{$\sim$}}}\hbox{$<$}}}}
\def\gtrsim{\mathrel{\hbox{\rlap{\hbox{\lower0.45em\hbox{$\sim$}}}\hbox{$>$}}}}
\def\gae{\mathrel{\hbox{\rlap{\hbox{\lower0.45em\hbox{$\sim$}}}\hbox{$>$}}}}
\def\lae{\mathrel{\hbox{\rlap{\hbox{\lower0.45em\hbox{$\sim$}}}\hbox{$<$}}}}

\title{
Universal equations and constants of turbulent motion}

\author{Helmut Z.\ Baumert}
\affil{Institute for Applied Marine and Limnic Studies -- IAMARIS e.V.\\ Bei den M\"uhren 69 A,
20457 Hamburg, Germany}

\begin{abstract}
\noindent
This paper presents a  parameter-free theory of shear-generated turbulence at asymptotically high Reynolds numbers
in incompressible fluids. It is based on a two-fluids concept. 
Both components are materially identical and inviscid. The first component is an ensemble 
of quasi-rigid dipole-vortex tubes (vortex filaments, excitations) as quasiparticles in chaotic motion. 
The second is a superfluid performing evasive motions between the tubes. 
The local dipole motions follow Helmholtz' law. The vortex radii scale with the energy-containing length scale. 
Collisions between quasiparticles lead either to annihilation (likewise rotation, turbulent dissipation) or to scattering 
(counterrotation, turbulent diffusion). There are analogies with birth and death processes of population dynamics and
their master equations and with Landau's two-flluid theory of liequid Helium. For free homogeneous decay the theory predicts the TKE to follow $t^{-1}$. With an adiabatic 
wall condition it predicts the logarithmic law with von K{\'a}rm{\'a}n's constant as $1/\sqrt{2\,\pi}= 0.399$. 
Likewise rotating couples form dissipative patches almost at rest ($\rightarrow$ intermittency) wherein under local quasi-steady conditions the spectrum 
evolves into an ``Apollonian gear'' as discussed first by \cite{herrmann90}. Dissipation happens exclusively at scale zero and at finite scales this system is frictionless and reminds of 
Prigogine's (1947) law of minimum (here: zero) entropy production. The theory predicts further the prefactor of the 3D-wavenumber spectrum (a Kolmogorov constant) as $\frac{1}{3}(4\,\pi)^{2/3}=1.802$, well within the scatter range of observational, experimental and DNS results.

\vspace{0.25cm}

\begin{flushright}
\textit{The diversity of problems in turbulence should not obscure\\
the fact that the heart of the subject belongs to physics.}\\ -- Falkowski and Sreenivasan (2006)\nocite{falkovichsreenivasan06}.
\end{flushright}

\end{abstract}


\section{Introduction\label{intro}}  
          
\noindent          
Many efforts to solve the turbulence problem rest on the idea that 
the Navier-Stokes equation (NSE) plays the role of a God equation and 
the application of a certain number of mathematical operations onto NSE could do it. 
In particular, the Fridman-Keller [1924] \nocite{kellerfridman24} series expansion of NSE played a prominent role, its zero$^{th}$  element being the Reynolds [1895] equation and the higher 
expansion elements subject to various closure hypotheses \cite[for a review see][]{wilcox06}. 
This assumption is also reflected in one of the Millenium-Prize problems of the Clay Mathematical Institute announced in 2000. However, until today these efforts could not answer most elementary
questions about turbulence. Here we explain why\footnote{Everywhere the pronomen ``we'' 
is used in this text, it means the two of us, the dear reader and the author.}.

A methodical alternative was chosen by \cite{prandtl26} who discussed turbulence in terms of 
analogies with molecular diffusion, gas kinetics, and Brownian motion in the interpretation by 
\cite{einstein05a}. Prandtl related his mixing length (\textit{Mischungsweg}) with the mean-free 
path of kinetic gas theory. This concept became popular but detailed questions could not be answered without use of measurements. Although Prandtl has been heavily criticized by 
\cite{batchelor53}, also other scientists derived free-hand analogies for turbulence, e.g. the early 
$K$-$\omega$ model by \cite{K42}, corrected and improved by \cite{saffman70},  further 
improved by \cite{wilcox06}. They are today part of a larger set of so-called two-equation 
turbulence models like $K$-$\varepsilon$, $K$-$K{\cal L}$, $K$-$\tau$ etc.\footnote{Comprehensive 
overviews on the history of turbulence theory can be found in 
\cite{davidson04} and \cite{davidsonetal11}}.

Three years before Prandtl, \cite{debeyehueckel1923} based their theory of electrolytes 
on the assumption that each ion is surrounded by a spherical "cloud" (or screen) of ions of 
opposite charge. 

The ideas of Prandtl and Debye and H\"uckel were early examples of 
constructive theories for large many-particle ensembles in very different
branches of physics, where knowledge of parts and pieces does not suffice 
to describe the system's global behavior. 

A similar methodical challenge was met later in the fields of superconductivity and 
superfluidity by groups around Landau and Feynman 
\cite[see][]{landaulifschitz_de83,landaulifshitz_eng87,feynman55,anderson2011}.
They realized that their fundamental equations and principles\footnote{Schr\"oderinger equation, Pauli principle etc.} hold perfectly for the single atom, 
but properties like electric conductivity (or the gloss of metallic surfaces) \textit{emerge} only 
when a larger \textit{ensemble} of atoms is put together so that new specific features and interrelations come into play, 
without violating the fundamental equations and principles, of course. 

For superfluidity it was necessary to supplement (or even replace) 
 fundamental equations by the two-fluid theory of Landau (1941) -- 
a not too far relative of the ion-cloud concept of \cite{debeyehueckel1923}. 

In many cases the formation of ensembles means to break symmetries. 
So  eventually the concept of \textit{emergence} and \textit{broken symmetries} 
gave physics in the last years a new and much less reductionistic face than it had 
until 1923 \cite[][]{glansdorffprigogine71,haken78,laughlin05,anderson2011}.

Encouraged by the victory of the a.\ m.\ two-fluid concept in superfluidity, Liepmann \cite[1961; loc. cit.][]{spiegel72} 
and later on \cite{spiegel72} made steps to test it for classical fluid turbulence.
Along this road, which we follow here too, \cite{spiegel72} was the very first who used 
terms like excitations, quasiparticles and two fluids in the context of classical fluid turbulence,
followed later by 
\cite{spalding85}\footnote{see \texttt{www.archive.org/details/nasa{/\_}techdoc{/\_}19860008217}}. 
These authors were at least fully aware of the principle, 
deep-rooted and almost philosophical problems with reductionistically NSE-based approaches. 
Ed Spiegel underlined his thorough conviction again just recently in the closing remarks of the report  
on the final session of the
 \textit{Turbulence Colloquium Marseilles 2011}\footnote{See \texttt{http://turbulence.ens.fr/}}. 
Besides ancient precursors 
of our ideas like Ren{\'e} Descartes (1596 -- 1650), who spoke about ``tourbillons'' 
forming the universe, and Lord Kelvin, who coined 1867  the notion ``vortex atoms'' 
\cite[loc.\ cit.][]{saffman92} we have to mention \cite{marmanis98} who possibly was 
the first to propose vortex dipoles as $the$ fundamental quasiparticles of turbulence.
Finally, the numerical Monte-Carlo eddy-collision methods with various vortex-filament primitives 
indicate that ideas developed by practitioners are not too far away from our 
theoretical views \cite[e.g.][]{andeme08}.

Below we elaborate the two-fluid concept in greater detail; not exhaustive because the
number of potential applications and side-problems is huge. But we show that, in an 
idealised sense, turbulence in an incompressible fluid at $Re\rightarrow\infty$ 
can be understood as a statistical 
many-body ensemble -- a tangle of vortex-dipole tubes (or filaments) taken as interacting \textit{discrete particles}. We will answer a number of open questions of turbulence without use of empirical parameters.

\section{Broken symmetry and irreversibility\label{symm}}

\noindent
In his anti-reductionistic article summarizing experiences of condensed-matter physics,
\cite{anderson72} states that ``more is different''. 
The turbulence theory presented below is an extreme example.
While the case of \textit{one} quasiparticle (an almost frictionless vortex dipole 
of zero net circulation) in a volume is non-dissipative and symmetric with respect 
to time and circulation, already the presence of \textit{two} quasiparticles 
in a common volume has the potential to break the symmetry and to give rise to the emergence
of turbulent dissipation: according to the laws of Helmholtz, dipoles are always in motion, may thus 
collide and may -- depending on the occasional collision angle -- 
form likewise-rotating and thus unstable couples. Their centers of mass stay almost at rest 
and evolve into a spectrum of smaller and smaller vortices until their kinetic energy is 
converted into heat at scales of size zero. In a sense we have here a most simple many-body 
problem because already the transition from ${\cal N}=1$ to ${\cal N}=2$ suffices to break the 
symmetry and allow the \textit{emergence} of dissipation.


Setting $Re\rightarrow\infty$ in NSE means vanishing viscosity such that only the Euler equation remains. However, the latter has non-unique solutions, visible already in the trivial case of the
inviscid Burgers equation. Additional information is needed to achieve uniqueness. 
I.e.\ with $Re\rightarrow\infty$ we first delete physics
(viscosity) in NSE to be forced then to \textit{add} (from outside) reasonable physics 
to finally reinstate uniqueness. But this is not all to be overcome. In addition, we have the problem of \textit{localization} and integrability of a solution. Classical weak vortex solutions of the Euler equation extend into the full volume and are thus not integrable. 
But real-world vortex ensembles exhibit individually localized vortices and finite scales. 

Those contradictions could only be resolved introducing quasiparticles embedded in
the two-fluid concept initiated earlier by Landau, Tisza, Feynman, Liepmann, Spiegel and Spalding, 
resting all on \cite{debeyehueckel1923}. 

\section{Vortex tubes and dipoles: Batchelor\\ \hspace{0.4cm}  couples and von-{K\'arm\'an} couples \label{couples}}

\noindent
{\bf Potential vortex.} 
The classical potential vortex around a closed vortex line (e.g.\ the centerline of a smoke ring) 
represents an exact weak solution of the Euler equation. The vortex line has infinitely thin diameter, 
infinitely high angular velocity, but finite circulation, $c=\pi\,\rho^2\,\omega< \infty$, 
where $\rho$ and $\omega$ are radial coordinate and vorticity.
The fluid outside the vortex line is inviscid and irrotational and the radial velocity
$v = \omega\,\rho \sim 1/\rho$ decreases with distance  $\rho$ from the centerline and extends
into space. It is not applicable here, as well as his relative, the Rankine vortex, because
they both are filling the space. 

{\bf Vortex tube. } Instead of the above we use the vortex-tube concept, which is a Rankine vortex
without sticking condition between its forced and free parts. I.e.\ its radial velocity
increases (like in a rigid vortex) linearly from center to tube radius $r$, and for $\rho > r$ there is 
an inviscid and independent potential flow governed by volume conservation. Thus vorticity is confined to the 
interior of a spaghetti-like tube as described e.\ g.\ by \cite{lugt79} and in greater detail by
\cite{pullinsaffman98} who quote papers by Kuo \& Corrsin, 1972, and Brown \& Roshko, 1974, showing tubes 
as dominating characteristic structures. A more recent simulation study has been presented by \cite{wilczek11} 
in form of a number of instructive movies on vortex ensembles in motion on the 
internet\footnote{\texttt{http://pauli.uni-muenster.de/tp/menu/forschen/}
\hspace{1cm}\texttt{ag-friedrich/mitarbeiter/wilczek-michael.html}}. 

The centerlines of our vortex tubes form either closed loops or they are attached to boundaries. 
The problem of stability of the tubes is discussed further below.

{\bf Batchelor couple.} 
This is a vortex \textit{dipole} made up of 
two anti-parallel vortices \cite[][]{lesieur97}
and symbolized by $(+\Uparrow-)$ or $(-\Uparrow+)$ 
where plus and minus mean the signs of vorticity within the vortices and the arrow 
the direction of motion of the couple or dipole. In classical
interpretation the flow field of one vortex moves the 
other vortex and \textit{vice versa}. 
The total circulation of a dipole is zero because the vorticities carry
opposite signs. In practice such couples are stable over moderate 
propagation times. In our idealized image they are made up of vortex
tubes, move frictionless with local center-of-gravity velocity $u=\omega\,r$ 
and conserve all their properties excepting their position in space because 
they propagate with their local $u$ into the same direction as
the fluid between the two tubes. In a dense ensemble of Batchelor couples 
their trajectories are no longer straight lines due to mutual interactions.
A  couple's kinetic energy density is $u^2/2=r^2\omega^2/2$. 

{\bf Von-K\'{a}rm\'{a}n couple.} 
It is a counterpart of the Batchelor couple, made up of likewise
rotating vortex tubes and symbolized by ($+\;\|\;+$) or ($-\;\|\;-$). 
Its total circulation does not vanish. Such a couple is known since long 
to be fundamentally unstable. Its kinetic energy is eventually dissipated 
into heat \cite[e.g.][]{lamb32}. Further below we discuss details of this process.

\section{Dipole chemistry in reaction-diffusion\\ \hspace{0.4cm}  approximation\label{rd}}

\noindent
For 2D trajectories it has been found by \cite{aref83} and \cite{eckhardtaref88} that the trajectories
are chaotic so that for very high $Re$ and 3D motions of tube-like vortex dipoles in form of a dense 
3D tangle we may assume also chaotic motions where collisions cannot be excluded. 
Let us consider an asymptotically high tube density so that the chaotic trajectories between 
collisions are short and locally homogeneous (in a statistical sense), like e.\ g.\ in the theory 
of Brownian motion. Then the total dynamic process of the tangle may be described as 
\cite[]{frisch95} ``\dots conservative dynamics punctuated by dissipative events\dots'':
some ``elastic'' collisions lead to energy-conserving reorganizations of Batchelor couples,
resembling turbulent diffusion, whereas other collisions are dissipative or ``inelastic'' and
lead to the formation of fundamentally unstable von-K\'{a}rm\'{a}n couples whose energy 
moves to smaller and smaller scales where it decays, resembling turbulent dissipation 
or dipole annihilation. 

\begin{figure}[htb]
\centerline{\includegraphics[width=8cm,height=7.5cm,keepaspectratio]{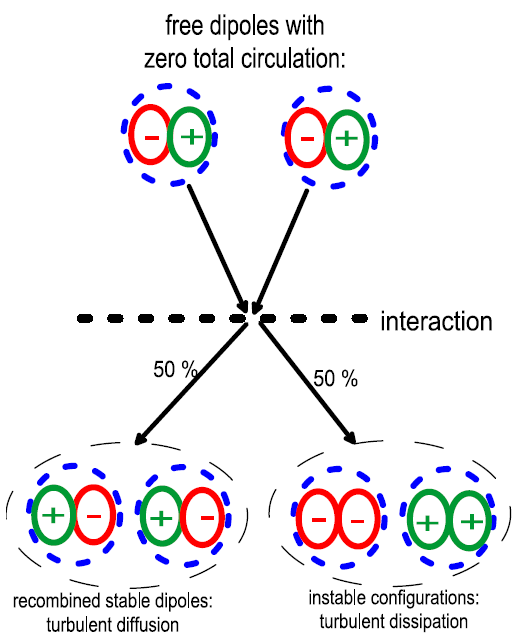}}
  \caption{Collisions of two dipoles: 
the left pathway is ``diffusive''; it is a recombination of dipole 
elements and chaotically scatters the trajectories (turbulent diffusion). The right pathway is 
``dissipative''; it evolves into an unstable vortex configuration which decays 
``somehow'' into heat. 
\label{f:gr1}}
\end{figure}
Figure \ref{f:gr1} shows the two potential results of a dipole-dipole collision.
For symmetry reasons the two pathways have identical probabilities of $\onehalf$. 

The simplest mathematical structure describing the statistics of the above two irreversible 
processes of (turbulent) diffusion and (turbulent) 
dissipation of dipoles is given by the following reaction-diffusion equation, 
which may be understood also as a special case of the Oregonator \cite[see e.g.\ Ch.\ 9 in][]{haken78}: 
\begin{eqnarray}
\label{n-1}
	\frac{\partial {\cal N}}{\partial t} + \frac{\partial}{\partial \vec x}
	\left( \vec U{\cal N}-\nu\, \frac{\partial {\cal N}}{\partial \vec x} \right) &=& {\cal F} - \beta\,{\cal N}^2 \,.
\end{eqnarray}
Here ${\cal N}$ is the volume density of dipole tubes (or filaments), $\nu$ is turbulent diffusivity, $\beta$ is a constant, 
 $\cal F$ is a source term describing the generation of dipole tubes and set equal to zero for the moment.
$\nu$ and $\beta$ are unknown so far. We only know for sure that the exponent of $\cal N$ 
in the last term of (\ref{n-1}) is really 2 because it needs \textit{two} colliding dipoles to 
generate (with probability $\onehalf$) unstable configurations which dissipate energy. 
But if energy is gone, a whole dipole is gone as our dipole tubes as quasiparticles 
differ from their inviscid potential-flow environment only by their kinetic energy. 

The presence of a mean flow and the corresponding advection of dipoles with the flow is sketched 
for reasons of completeness by the term with mean-flow vector $\vec U$. For simplicity
we set further always $\vec U \equiv 0$.

One may look at (\ref{n-1}) only from the viewpoint of pure analogy with
chemical reaction-diffusion processes. But one may also derive this equation over many pages
from scratch, beginning with a master equation for the probabilities as sketched earlier \cite[Ch. 5.6]{baumert05b}. 
Fortunately, this has already been done by other authors many years ago for whole classes of such processes,  
and went into the many textbooks on stochastic-dynamic systems, physical kinetics and other fields 
\cite[e.g.][]{kraichnan68,haken78,haken1983,stratonovich92,stratonovich94}.
Therefore we will not repeat here these formal steps and refer 
the more technically interested reader to the literature mentioned.

\section{Two fluids, dressed and naked tubes\label{twofluids}}

\noindent
{\bf Two fluids.} We consider a circulation-free volume filled with two different
forms of an materially identical incompressible fluid: 
\begin{itemize}
  \item [(a)] inviscid, quasi-rigid and incompressible but deformable vortex tubes as quasiparticles with 
	finite radius $r$ and vorticity module $\omega$, vorticity bundled within the tube; 
  in our analogy with the kinetic theory of gases a dipole tube is an analogue of a particle,
  \item [(b)] the inviscid fluid \textit{between} the dipole tubes; it is and in the analogous kinetic 
	theory of gases an analog of the vacuum.
\end {itemize}
Fluid (b) behaves like a super-fluid and receives no force 
from the moving quasi-rigid dipole tubes (d'Alambert's paradox). 
The fluids differ only in their state of motion. While the tubes rotate
around their (in general curvilinear) axes and move locally relative to the volume
according to Helmholtz' laws, the fluid between the tubes
performs corresponding evasive motions according to the 
principle of volume conservation.

{\bf Naked tubes.} 
Above we have used the concept of vortex dipoles made up of ideal vortex tubes
immersed in an inviscid fluid and exchanging no energy with it. However, this
concept is only applicable if the vortex is a rigid body. If it is a fluid, the problem
of stability arises because the quasi-rigid vortex tube as an exact solution of 
the Euler equation is accompanied by the following pressure head as a 
consequence  of inertial (centrifugal) forces:
\begin{equation}\label{head}
p = p_0+ \frac{\rho }{2} \times \omega^2r^2.
\end{equation}
Here $p_0$ is the background pressure of a laminar reference flow, 
e.g.\ in the ocean the depth-depending hydrostatic pressure.
If the pressure outside the vortex would simply be $p_0$ then, 
due to the action of the outwards-directed pressure force given by the 
second term in (\ref{head}), the vortex would loose stability against small
perturbations. However, tubes of finite radius are observed to be during their 
finite life time relatively stable in real-world turbulence.

{\bf Dressed tubes.} 
This contradiction can be explained by the consideration of ensemble effects.
It has been noted in the Introduction of this paper that in many-body problems 
like turbulence the phenomenon of \textit{emergence} deserves special attention.
This means that in a (local) volume element with a larger number (${\cal N} \gg 1$) of similar vortex 
tubes (a local ``cloud'') the tube ensemble itself generates the background pressure (\ref{head}) 
which keeps eventually all the individual tubes -- at least in the center of the cloud --
in a ``sufficiently stable'' state. 

In thermodynamically open systems like turbulence such a cloud represents 
a quasi-steady state \cite[\textit{Flie\ss gleichgewicht} in the sense of ][]
{bertalanffy53,glansdorffprigogine71,haken78,haken1983}.
I.e.\ the processes of dipole generation (see below) and their annihilation by collisions 
almost compensate each other. Any quasiparticle (dipole) has thus only a limited 
statistical life time. Therefore ``sufficient stabilization'' of a dressed dipole tube by
a cloud means to guaranty stability  only in a statistical sense during its (statistical) life time.

\section{Particle number, TKE,\\ \hspace{0.4cm}   and r.m.s.\ vorticity frequency \label{tke}}

\noindent
Consider a small volume element $\delta V$ populated by an ensemble of $j = 1 \dots {\cal N}$ 
dipoles with individual effective vortex radii, $r_j$, and r.m.s.\ vorticity moduli, $\omega_j$. 
The latter can be interpreted as r.m.s.\ values of individual dipoles $j$ as follows,
\begin{eqnarray}
	\omega_j^2 &=& \frac{1}{2}\left[ (- \hat \omega_j)^2 + (+ \hat \omega_j)^2\right]  \;=\;\hat \omega_j^2\,,
\end{eqnarray}
where $+ \hat \omega_j$ and $- \hat \omega_j$ are the individual vorticities of the two vortex tubes forming the dipole.

The dipole energy is conserved as long as dissipative events do not take place. 
The volume density of dipoles is ${\cal N}/\delta\cal V$.
The total TKE within $\delta \cal V$ is the sum of the kinetic energies
of the individual dipoles:
\begin{eqnarray}
	K_{\delta \cal V}&=& 
	\frac{1}{\delta\cal V}
	\sum_{j\,\in \,\delta\cal V} \,
	\frac{1}{2}\,r_j^2\,\omega_j^2\;
	=\;\frac{\cal N}{\delta \cal V}\;\bar k\;\label{Knk-1},\\
	\omega_{\delta \cal V}^2 &=& \frac{1}{\delta\cal V}\sum_{j\,\in \,\delta\cal V}\omega_j^2
	\;=\;\frac{{\cal N}}{\delta \cal V}\;\bar\omega^2 \,.\label{omeganom-1}
\end{eqnarray}
Multiplication of (\ref{Knk-1}, \ref{omeganom-1}) with $\delta\cal V$ gives 
\begin{eqnarray}
	K\;=\; \delta \cal V\cdot K_{\delta \cal V}&=& 
	\sum_{j\,\in \,\delta\cal V} \,
	\frac{1}{2}\,r_j^2\,\omega_j^2\;
	=\;{\cal N}\;\bar k\;\label{Knk-2},\\
	\omega^2 \;=\; \delta {\cal V}\cdot 
	\omega^2_{\delta \cal V} &=& \sum_{j\,\in \,\delta\cal V}\omega_j^2
	\;=\;{\cal N}\;\bar\omega^2\,.\label{omeganom-2}
\end{eqnarray}
$K_{\delta \cal V}$, $\omega_{\delta \cal V}$ are local volume densities 
of TKE and r.m.s.\ vorticity magnitude, respectively. They and  $K, \omega$ 
are extensive variables by definition, i.e.\ they scale with the dipole number in 
$\delta\cal V$. $\bar k$ and $\bar \omega$ are ensemble averages and as such 
intensive variables which do not change when new particles with 
average properties are added to $\delta \cal V$:
\begin{eqnarray}
	\bar k&=& \frac{1}{{\cal N}}
	\sum_{j\,\in \,\delta\cal V} \label{Knk-3}
	\,\frac{1}{2}\,r_j^2\,\omega_j^2\,,\\
	\bar \omega^2&=& \frac{1}{{\cal N}} \sum_{j\,\in \,\delta\cal V}\omega_j^2\,.\label{omeganom-3}
\end{eqnarray}
We turn now to equation (\ref{n-1}) where, according to  (\ref{Knk-2}, \ref{omeganom-2})
we replace $\cal N$ with $K/\bar k$ and $\omega/\bar\omega$, respectively, to get eventually
balance equations for the extensive variables $K$ and $\Omega=\omega/2\pi$, provided that the intensive
variables $\bar k$ and $\bar \omega$ vary sufficiently weakly in time and space compared 
with the dipole number $\cal N$:
\begin{eqnarray}
\label{K-1}
	\frac{\partial K}{\partial t} +\frac{\partial}{\partial \vec x}
	\left( \vec U\,K - \nu\, \frac{\partial K}{\partial \vec x} \right)
	&=&  {\cal F}_K-\beta_K	\,K^2 ,\\
\label{Om-1}
	\frac{\partial \Omega}{\partial t}+ \frac{\partial}{\partial \vec x}
	\left( \vec U\,\Omega\,-\nu\,\frac{\partial\Omega}{\partial\vec x}\right)
	&=&{\cal F}_{\Omega} -\;\beta_{\Omega}\,\Omega^2\,.
\end{eqnarray}
Here $\Omega=1/T$ is the ordinary\footnote{While $\omega=2\pi/T$ is an angular 
frequency, $\Omega=1/T$ is an ordinary frequency. } vorticity frequency and related with $\omega$ by
the often used constant $\kappa$ which further below appears to be von-K\'{a}rm\'{a}n's constant:
\begin{eqnarray}
	\omega &=& 2\,\pi\,\Omega \;=\;\Omega/\kappa^2\,\label{Omega},\\
	\kappa &=& (2\,\pi)^{-1/2}\approx \,0.399 \,\label{kappa1}\,.
\end{eqnarray}
We see that 
\begin{eqnarray}
	\beta_K &=& \beta/\bar k\,,\quad\quad\quad \beta_{\Omega} \;\;=\;\; 2\,\pi\, \beta/\bar\omega\,,\\
	{\cal F}_K &=& \bar k\, {\cal F}\,,\quad\quad\quad {\cal F}_{\Omega} \;\;=\;\; \bar \omega\, {\cal F}/2\,\pi  \,\label{kappa1}\,.
\end{eqnarray}

\section{Mixing length and eddy viscosity \label{nu}}

\noindent
\cite{prandtl25} hypothesized that his mixing length has a two-fold meaning. 
It should be ``considered as the diameter of the masses of fluid moving as a whole'' 
(in our picture: as a characteristic vortex radius) 
or ``as the distance traversed by a mass of this type before it becomes blended 
	in with neighboring masses\dots'' (in our picture: as a mean free path).
\cite[loc.\ cit.][]{bradshaw74}. I.e.\ he stated a tight relation between the 
radius and the free path of a dipole in a dense ensemble. Indeed, for a 
dipole tangle this can be shown. We first define the ensemble average,
$\bar r$, of the vortex radii as an average over a volume element weighted by vorticity as follows,
\begin{equation}\label{r2}
	\bar r^2 \;\;=\;\; \frac{\sum\;r_j^2\,\omega_j^2}
	{\sum \omega_j^2} \;=\; \frac{2\,K}{\omega^2}\;=\; \frac{2\,K}{(2\,\pi\,\Omega)^2}\,.
\end{equation}
We now determine the eddy viscosity, $\nu$,  
in analogy to the theory of Brownian motion \cite[][]{einstein05a} in terms of a mean free path, 
$\lambda$, and a mean free flight time, $\tau =\lambda/\bar u$:
\begin{eqnarray}
	\nu &=& {\lambda^2}/{2 \,\tau}\,\label{nu-1}\,.
\end{eqnarray}
Here $\bar u$ is the mean velocity of a dipole,
\begin{eqnarray}
	\bar u^2 &=& 2\,K \;\;=\;\; \sum\,u_j^2 \;\;=\;\; \sum\,r_j^2\,\omega_j^2\,. 
\end{eqnarray}
\begin{figure}[tbp] 
\centerline{\includegraphics[width=6cm,height=5.81cm,keepaspectratio]{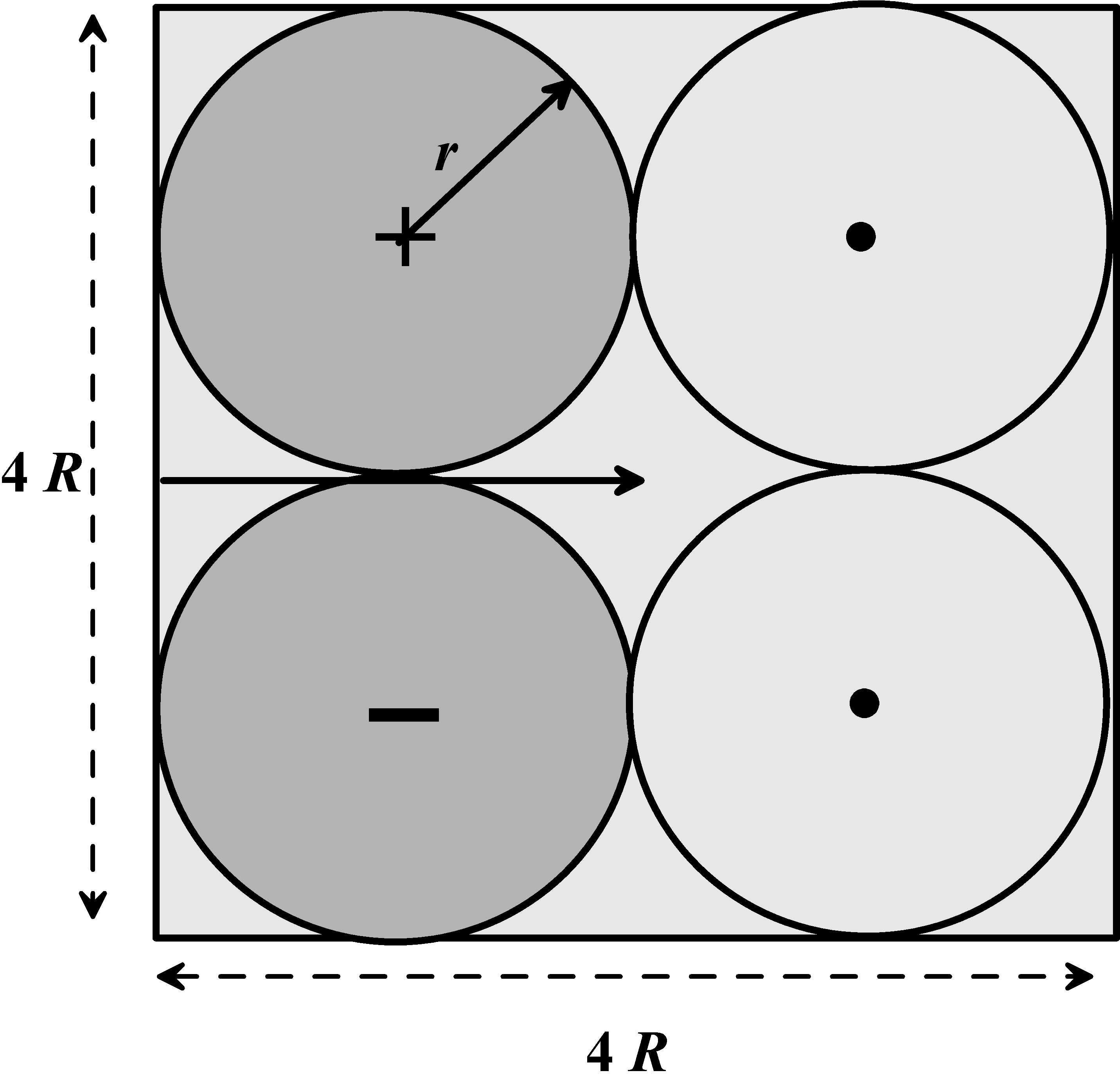}}
  \caption{Local cross section through a dipole tangle  far from boundaries
at maximum dipole density ($Re\rightarrow\infty$). 
The dark grey dipole in the left half square labels the initial 
situation, the right half square a situation after a translation motion from left to right 
into its collision position, where it is either annihilated or scattered. A free dipole
obviously occupies statistically a cross sectional area of $(4\,\bar r)^2$.}
\label{f:vorticons-area-1}
\end{figure}
For a turbulent tangle of dipole tubes at $Re=\infty$ in a quasi-steady  state far from solid boundaries 
the mean free path will clearly be short but cannot vanish because there is an equilibrium of dipole
generation, annihilation, and motions that lead to the latter. This implies that 
\cite[][]{baumert05b} \begin{eqnarray}\lambda &=& 2\,\bar r\,\label{lambda-2}.\end{eqnarray}
Algebra finally gives the following: 
\begin{eqnarray}
	\nu &=& K/\pi \, \Omega \,.\label{nu-2}
\end{eqnarray}
This relation is called the Prandtl-Kolmogorov relation. 

\section{Parameters \begin{boldmath}$\beta_K$, $\beta_{\Omega}$, and $\zeta$ \end{boldmath}\label{params}}

\noindent
We consider an initial-value problem (${\cal F}_K=0$, ${\cal F}_{\Omega}=0$) in a spatially homogeneous 
volume where $\partial K/\partial \vec x=0$ and $\partial \Omega/\partial \vec x=0$ 
such that equations (\ref {K-1}, \ref{Om-1}) reduce to
\begin{eqnarray}
	\label{tke-0}
	\frac{dK}{dt}+\beta_K\,K^2&=& 0\;,\\
	\label{om-0}
	\frac{d\Omega}{dt}+\beta_{\Omega}\,\Omega^2&=& 0\;.
\end{eqnarray}
For large $t$ follows that
\begin{eqnarray}
\label{tke-1}
	K(t)& = & (\beta_K\,t)^{-1}\;,\\
\label{om-1}
	\Omega(t)&=& (\beta_{\Omega}\,t)^{-1}\;,
\end{eqnarray}
and, according to (\ref{nu-2}), for eddy viscosity holds 
\begin{eqnarray}
\label{nu-3}
	\nu(t) &=& \frac{K(t)}{\pi\,\Omega(t)}
	\;=\;\frac{\beta_{\Omega}}{\pi\,\beta_K}\;=\;\mbox{const.}
\end{eqnarray}
Equation (\ref{tke-1}) coincides with the results of a fairly general 
similarity analyses of the Navier-Stokes equations by \cite{oberlack02}
and with experimental results by \cite{dickeymellor80}.

We now use the definition of the dissipation rate, $\varepsilon$, which is
the last term on  the right-hand side of (\ref {K-1}) and carries the units of
of TKE (density) per time, i.e.\ $\varepsilon\sim K/T\sim K\,\Omega$, [m$^2$ s$^{-3}$]. 
For reasons of convenience we write this variable as follows and introduce a still unknown
auxiliary variable, $\zeta$: 
\begin{eqnarray}
	\varepsilon &=& \beta_K\,K^2\;\;=\;\; \zeta \,\frac{K\,\Omega}{\pi}
	\;=\; \zeta \,\frac{K^2}{\nu\,\pi^2}
	\,\label{eps1},
\end{eqnarray}
where we made use of (\ref{nu-2}). This implies 
\begin{eqnarray}
	\beta_K &=& {\zeta}/{\nu\,\pi^2} \,\label{eps11}.
\end{eqnarray}
When the dipoles behave in general like in free decay
then we may use (\ref{nu-3}) which gives then the expressions
\begin{eqnarray}
	\beta_{\Omega} &=& \zeta/\pi \,\label{betaom1},\\
	\beta_K &=& \zeta/\nu \,\pi^2 \label{betak1}.
\end{eqnarray}
With these results we may rewrite (\ref {K-1}, \ref{Om-1}) as follows:
\begin{eqnarray}
\label{K-11}
	\frac{\partial K}{\partial t} +\frac{\partial}{\partial \vec x}
	\left( \vec U\,K - \nu \, \frac{\partial K}{\partial \vec x} \right)
	&=&  {\cal F}_K-\frac{\zeta\,K^2}{\nu\,\pi^2}\, ,\\
\label{Om-11}
	\frac{\partial \Omega}{\partial t}+ \frac{\partial}{\partial \vec x}
	\left( \vec U\,\Omega\,-\nu \,\frac{\partial\Omega}{\partial\vec x}\right)
	&=&{\cal F}_{\Omega} -\,\frac{\zeta}{\pi} \,\Omega^2\,,
\end{eqnarray}
with $\nu$ given by (\ref{nu-2}) and $\zeta \equiv 1$ as shown further below.

\section{TKE and vorticity generation\\ \hspace{0.42cm}   by mean-flow shear\label{generation}}

\noindent
{\bf Generation of TKE.} 
To further complete equations (\ref{K-11}, \ref{Om-11}), we have, besides $\zeta$, to specify the 
source terms ${\cal F}_K, {\cal F}_{\Omega}$. The specification of ${\cal F}_K$ for shear flows is trivial 
because it is given by the classical losses of the mean-flow energy balance and can therefore be copied from
textbooks like \cite{wilcox06} or \cite{schlichting-gersten00}:
\begin{eqnarray}\label{tke-production}\nonumber
	{\cal F}_K &=& 	-\sum_{i,j=1}^3 \langle u_i'\,u_j'\rangle\, \frac{\partial U_i}{\partial x_j}\\
	&=& \nu \, S^2 - \frac{2}{3}\,K\,\sum_{i,j=1}^3 \delta_{ij}\,\frac{\partial U_i}{\partial x_j}\,,
\end{eqnarray}
where $	-\langle u_i'\,u_j'\rangle $ is the Reynolds stress tensor defined as 
\begin{equation}\label{downhill-1}
	-\langle u_i'\,u_j'\rangle  = 2\,\nu \, S_{ij} -\frac{2}{3}\,\delta_{ij}\cdot K
\end{equation}
and $S_{ij}$ is the rate of strain tensor defined as 
\begin{equation}\label{downhill-2}
	S_{ij} = \frac{1}{2}\,\left( \frac{\partial U_i}{\partial x_j}+\frac{\partial U_j}{\partial x_i}\right)\,.
\end{equation}
$S^2$ is the total instantaneous shear squared,
\begin{equation}\label{def-shear}
	S^2 \; = \; \sum_{i,j=1}^3 \left( 
	\frac{\partial U_j}{\partial x_i}+
	\frac{\partial U_i}{\partial x_j} \right)\;
	\frac{\partial U_i}{\partial x_j}\,,
\end{equation}
$U_i$ is the $i^{th}$ component of the mean flow velocity vector $\vec U= (U_1, U_2, U_3)^T$ 
and $\delta_{ij}$ is the Kronecker symbol which is zero for $i \ne j$ and unity for $i=j$. 
In the simple case of a plane horizontal flow with vertical shear like wind over plane terrain or flow 
in a plane wide channel we have $i=1$ and $j=3$ such that 
(\ref{tke-production}) reads as follows: 
\begin{eqnarray}\label{tke-production-1}
	{\cal F}_K &=& 	\nu \, S^2 \,.
\end{eqnarray}

{\bf Generation of vorticity.} We use a fundamental macroscopic argument by \cite{tennekes89}. 
It has first been cast into mathematical form by \cite{baumertPeters00, baumertPeters04}
and carefully discussed by \cite{kantha04,kantha05}, by  \cite{kanthaetal05} and by
 \cite{kanthaclayson07}. 
Tennekes hypothesized that, in a neutrally stratified homogeneous shear flow, an energy-containing turbulent length scale, 
$\cal L$, cannot, on dimensional grounds, depend on the ambient shear. 

Further, also on dimensional grounds we have ${L}\sim K^{1/2}\,\Omega^{-1}$ 
and thus ${ L}^2 \sim K\,\Omega^{-2}$ 
such that the time evolution of the length scale follows 
\begin{equation}\label{tennekes-1}
	\frac{1}{{ L} ^2}\,\frac{dL ^2}{dt}\;\sim \;\frac{1}{K} \,\frac{dK}{dt}-2\,\frac{1}{\Omega}\,\frac{d\Omega}{dt}\,.
\end{equation}
We replace $dK/dt$ with ${\cal F}_K$ according to (\ref{tke-production}) and find
\begin{equation}\label{tennekes-2}
		\frac{1}{{L} ^2}\,\frac{dL ^2}{dt}\;\sim \; \frac{S^2}{\pi\,\Omega} - 2\,\frac{1}{\Omega}\,\frac{d\Omega}{dt}\,.
\end{equation}
Tennekes' hypothesis means that the evolution of $L$ cannot be controlled by $S$ which means $d{L}^2/dt=0$ and implies that 
\begin{equation}\label{tennekes-3}
	\frac{S^2}{\pi\,\Omega} \;=\; \frac{2\,{\cal F}_{\Omega}}{\Omega}\,.
\end{equation}
where we replaced $d\Omega/dt$ with ${\cal F}_{\Omega}$.
Algebra gives
\begin{eqnarray}
	\label{def-omega-1}
		 {\cal F}_{\Omega} &=& {S^2}/{2\,\pi}\,
\end{eqnarray}
and for the simple case of a plane horizontal flow with vertical shear we may complete 
with some algebra equations (\ref{K-11}, \ref{Om-11}) as follows:
\begin{eqnarray}
\label{K-111}
	\frac{\partial K}{\partial t} +\frac{\partial}{\partial \vec x}\left( \vec U\,K - \nu \frac{\partial K}{\partial \vec x} \right)&=&  \nu\left(S^2-\zeta\Omega^2\right),\\
\label{Om-111}
	\frac{\partial \Omega}{\partial t}+ \frac{\partial}{\partial \vec x}\left( \vec U\,\Omega-\nu \frac{\partial\Omega}{\partial\vec x}\right)
&=&\frac{1}{\pi} \left( \frac{S^2}{2} -{\zeta}\Omega^2\right)
\end{eqnarray}
with $\nu$ again given by (\ref{nu-2}). 

\section{Turbulent boundary layer\label{boundlayer}}

\noindent
{\bf Boundary conditions.}
Consider a horizontally homogeneous flow and its stationary boundary layer close to a plane solid wall at $x_3 = z = 0$ where
for convenience $z$ is introduced here as the only relevant coordinate, pointing orthogonal from the wall into the fluid. Thus the shear is
\begin{equation}\label{newshear}
	S \;=\; \left|dU/dz\right|.
\end{equation}
This special case means that in (\ref{K-111}, \ref{Om-111}) the horizontal and time derivatives vanish and it remains
\begin{eqnarray}
\label{K-1111}
	-\frac{d}{dz}\left( \nu\, \frac{dK}{dz} \right)&=&  \nu\,\left(S^2-\zeta\,\Omega^2\right),\\
\label{Om-1111}
	-\frac{d}{dz}\left( \nu\, \frac{d\Omega}{dz}\right)
&=&\frac{1}{\pi} \left( \frac{S^2}{2} -{\zeta}\,\Omega^2\right)
\end{eqnarray}
The diffusive TKE flux into the viscous sublayer at $z=z_0$ has  to vanish
in the sense of an adiabatic boundary condition, 
\begin{equation}\label{K-25}
	\left( \nu \,dK/dz\right)_{z=z_0} \;=\; 0\,,
\end{equation}
such that also the flux divergence on the left-hand side of (\ref{K-1111}) vanishes, giving
$K=K_0$ and $\nu\,(S^2 - \zeta\,\Omega^2) =0$ or
\begin{equation}\label{K-26}
	\Omega  \;=\; S/\sqrt{\zeta}\,.
\end{equation}

{\bf Logarithmic law of the wall.} 
We insert (\ref{K-26}) into (\ref{Om-1111}). The unknown  $\zeta$ cancels 
out\footnote{The respective text in \texttt{http://arxiv.org/pdf/0907.0223.pdf} contains an algebraic error
but without later consequences, fortunately.}
and we have to solve the following equation for $S=S(z)$, 
\begin{equation}\label{Om-5}
	{2\,K_0}\,\frac{d}{dz}\left( \frac{1}{S}\,\frac{dS}{dz}\right)  \;=\; S^2\,,
\end{equation}
which gives
\begin{equation}\label{Om-6}
	S(z)\;=\; \frac{dU}{dz}\;=\; \frac{\sqrt{2\,K_0}}{z}\,.
\end{equation}
Integration of (\ref{Om-6}) gives the logarithmic law of the wall. 

In boundary layer theory the bottom shear stress is 
defined in terms of the squared friction velocity, $u_f^2$,
\begin{equation}\label{bss}
	u_f^2 \;=\; \nu\, \frac{dU}{dz}\;=\; \frac{K_0}{\pi \,\Omega}\,S\,,
\end{equation}
and with (\ref{K-26}) it follows that
\begin{equation}\label{k-47}
	K_0 \;=\; \pi \,u_f^2 / \sqrt{\zeta}\, .
\end{equation}
This allows to rewrite (\ref{Om-6}) as follows,
\begin{equation}\label{Om-7}
	\frac{dU}{dz}\;=\; \frac{u_f}{\tilde \kappa\,z}\,,
\end{equation}
with $\tilde\kappa$ as a modified von-K\'{a}rm\'{a}n constant 
defined with respect to (\ref{kappa1}) through
\begin{equation}\label{Om-8}
	\tilde \kappa \;=\; \kappa \, \zeta^{1/4} \,.
\end{equation}
Integration of (\ref{Om-7}) provides us with 
\begin{equation}\label{Om-9}
	U(z)\;=\; \frac{u_f}{\tilde \kappa}\, \ln\left(\frac{z}{z_0}\right).
\end{equation}

\begin{figure}[hbt]
\centerline{\includegraphics[width=6cm,height=7cm,keepaspectratio]{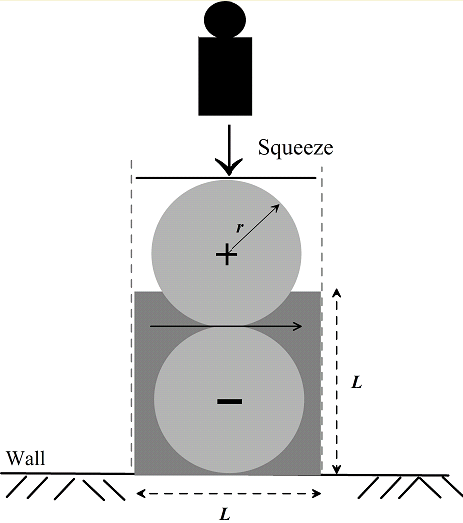}}
\caption{Cross section through a dipole sheet at a
solid boundary. In this Figure we have $r=\bar r$, i.e.\ the sketched 
quasiparticle is to be understood as an ensemble average. }
\label{f:vorticons-boundary}
\end{figure}

{\bf Mixing length $L$\label{mixinglength}.}
Consider the definition of the effective (statistically averaged) dipole radius of an ensemble through (\ref{r2}). We solve this
equation for $K$ and express the TKE in terms of $\bar r$ and $\Omega$ as follows:
\begin{equation}\label{10}
	K\;=\; 2\,\pi^2 \,\bar r^2\, \Omega^2\,.
\end{equation}
Following now Hinze [1959, p. 279, eq. 5-2], in present notation Prandtls 
mixing length $L$, which is termed also the ``energy-containing length scale'' in the literature, 
is defined in terms of eddy viscosity and shear as follows: 
\begin{equation}\label{11-a}
	\nu\;=\; {L}^2\,\left|\frac{dU}{dz}\right|\;=\;{L}^2\,S\,.
\end{equation}
Due to the eddy-viscosity formula (\ref{nu-2}) relation (\ref{11-a}) gives 
\begin{equation}\label{11-b}
	{L}^2\;=\; \frac{K}{\pi\,\Omega^2\,\zeta^{1/2}}\,,
\end{equation}
so that in the neighborhood of a solid wall we get with (\ref{K-26}) the following result:
\begin{equation}\label{11-c}
	K \;=\; \pi\, {L}^2\, \Omega^2\,\sqrt{\zeta}\,.
\end{equation}
Comparing (\ref{10}) with (\ref{11-c}) gives 
\begin{equation}\label{11-d}
	{L}\;=\; {\bar r}/{\tilde \kappa}.
\end{equation}
The physical meaning of $L$ can be understood as follows (see Fig.\ \ref{f:vorticons-boundary}).
If we may set $\zeta=1$ then (\ref{11-d}) gives together with (\ref{kappa1}) and (\ref{Om-8}) the following relation:
\begin{equation}
	{L}^2 = 2\times (\pi\,\bar r^2)\,.\label{12-a}
\end{equation} 
It means that $L$ is the length of a square with an area equal to the 
cross sectional area of a dipole (in a statistically averaged sense)
because $\pi \, \bar r^2$ is the cross-sectional area of one vortex tube.
In an asymptotic sense this case corresponds to the maximum
deformation of a dipole and justifies to set $\zeta \equiv 1$.

While Prandtl's mixing-length concept was applicable only in the vicinity of solid 
boundaries so that it attracted respectful criticism \cite{wilcox06} our concept is a generalization of Prandtl's concept and works 
also far from boundaries, even in the free stream of stratified fluids where $L$ may approach 
the Thorpe scale and/or the Ozmidov scale, depending on the conditions 
(Baumert and Peters 2004)\nocite{baumertpeters04}. 

We summarize this section as follows:
\begin{eqnarray}
	\nu&=&  u_f\,L\,\label{a-1}, \\
	L&=& \kappa\,z \,\label{a-1} ,\label{a-2}\\
	z &=& L/\kappa \;=\;\bar r/\kappa^2 \label{a-4}\,.
\end{eqnarray}

\section{Spectra, dissipative patches,\\ \hspace{0.65cm} and spectral constants}

\noindent
{\bf Spectra. \label{spectra}}
Up to now our discussion was concentrated on only a few scales like $\bar r$ and $L$ interrelated by $\kappa$,
and on $T$ interrelated with $\Omega$ and $\omega$ etc. However, the reality of turbulence exhibits a whole range of scales
at which fluctuations occur. They have relatively stable spectral properties. This problem has attracted  early attention by
\cite{richardson22} and \cite{taylor37}. \cite{kolmogorov41c} found on dimensional grounds that 
the kinetic energy spectrum as function of wavenumber, wherein energy flows steadily from the large (energy-containing) scales to the much smaller dissipative scales, may be written as follows,
\begin{equation}\label{kolmo1}
	d{ K}=\alpha _1 \,\varepsilon ^{\alpha _2}\,k^{-\alpha _3}\, dk\;.
\end{equation}
Here $k = 2\pi/\Lambda $ is the wave number and $\Lambda $ the wavelength. 
$\varepsilon $ is the dissipation rate of TKE.  Dimensional
arguments force that that $\alpha _2 = 2/3$ and $\alpha _3 = 5/3$, 
in agreement with the oceanographic observations by \cite{grantetal59}
in a tidal inlet with $Re \approx 10^8$ and a depth of about 100 m. 
A value for $\alpha_1$ is derived below.

{\bf Devil's gear.}
Our view of the above Kolmogorov-Richardson cascade has been filled with life through a 
numerical simulation study by \cite{herrmann90} who demonstrated that 
Kolmogorov's value for $\alpha_3$ corresponds numerically 
to the data  of a space-filling bearing \cite[see also][]{herrmannetal90}. 
The latter is the densest non-overlapping (Apollonian) circle packing in the plane, 
with side condition that the circles are pointwise in contact but able to rotate freely, 
without friction or slipping \cite[a ``devil's gear'' \textit{sensu}][]{poeppe04}.
The contact condition for two different ``wheels'' with indizes 1 and 2 of the gear  reads
\begin{equation}\label{contact1}
	u = \omega_1\;r_1 = \omega_2\;r_2\; ,
\end{equation}
where $u$ is necessarily constant throughout the gear 
and governed by the energy of the decaying (initially energy-containing)
vortex pair as $u=\sqrt{2\,K}$. It follows that 
\begin{equation}\label{contact2}
	\omega_2 = \omega_1\;\frac{r_1}{r_2}\; ,
\end{equation}
and for very small $r_2$ the frequency $\omega_2$ may become high, even acoustically relevant.

\begin{figure}[htb]
\centerline{\includegraphics[width=6cm,height=7.0cm,keepaspectratio]{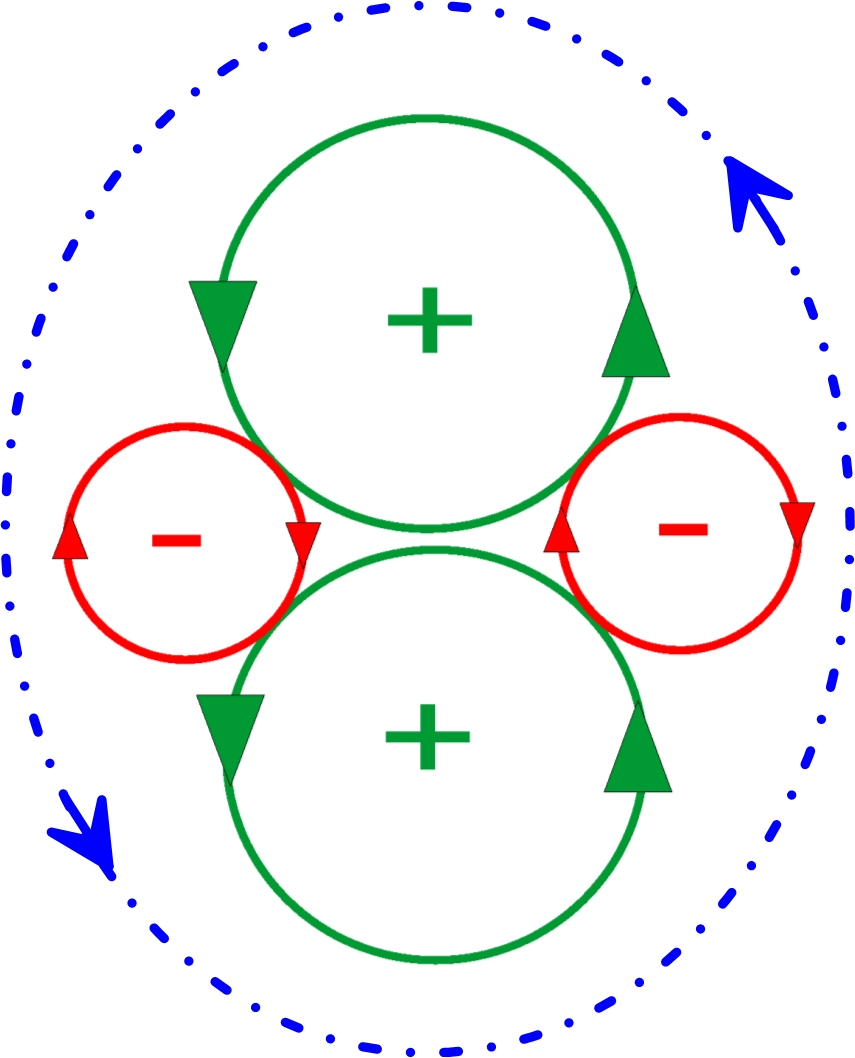}}
  \caption{Cross section through the first developmental stage of a dissipative patch, i.e.\ of an unstable pair  
of likewise rotating vortices resulting from a dipole-dipole collision (right pathway in Fig.\ \ref{f:gr1}). 
The green circles represent primary energy-containing vortices. They do \textit{not} touch each other 
due to spontaneously formed secondary vortices which initiatiate a whole vortex cascade. 
The broken blue line and the arrows symbolize  the slow rotation of the patch 
around its center of mass.  \label{f:gr2}}
\end{figure}

{\bf The dissipative patch.}
If the above gear is frictionless then the question arises where energy can be dissipated. In
a real fluid with non-zero viscosity, dissipation happens at all scales, mainly but not exclusively 
where the velocity gradient is highest, here: at a scale vanishing with $Re\rightarrow\infty$ 
to the size zero. Our dissipative patch 
(Fig.\  \ref{f:gr2}  shows 
the first stage of its formation) is thus ``almost frictionless'' and a Hamiltonian clockwork, excepting 
scales of size zero. 

The formation of a fully developed spectrum of ``wheels'' from Fig.\ \ref{f:gr2}  deserves 
certain perturbations ``from the sides'', a condition which is guaranteed by the random 
reconnection/recombination and scatter processes sketched in
the left half of  {Fig.\ \ref{f:gr1}} and also by the incomplete mutual 
pressure  compensation of the vortices in our vortex ensemble. 
In a quasi-steady state these perturbations initiate roll-up instabilities at
the boundaries of the respective larger vortices so that eventually and in a statistical sense 
a patch like in Fig.\ \ref{f:gr2} is formed and evolves steadily into a fully developed gear. 

Our Fig.\  \ref{f:gr1} illustrates the possible results of a dipole-dipole collision. While the left pathway shows 
the recombination of \textit{counter}-rotating vortices from counter-rotating vortices, the right shows the formation 
of a couple of \textit{likewise} rotating vortices from counter-rotating vortices. The latter then revolve around a 
common center of mass which remains nearly at rest (Fig.\ \ref{f:gr2}). Such a couple 
is fundamentally unstable \cite[]{sommerfeld48}, a quasi-steady dissipative patch evolving 
into a full gear in the sense of the mechanisms discussed in the last Chapter. 
This picture lets us expect that dissipation should exhibit a spatially \textit{patchy} 
behaviour which we may also call \textit{intermittency}. This problem has been studied extensively by various authors from other points of view \cite[see e.g.][]{frisch95} and cannot yet be discussed here 
from our viewpoint in greater detail. 

\begin{figure}[htb]
\centerline{\includegraphics[width=7.5cm,height=4cm,keepaspectratio]{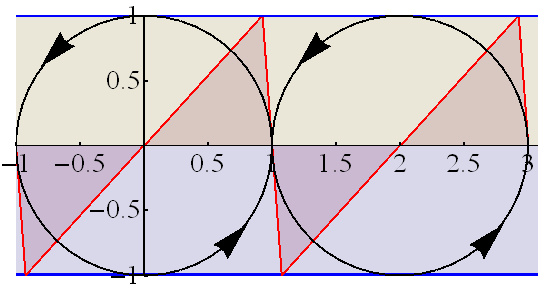}}
  \caption{Outer limits of a dissipative patch ($c.f.$ Fig.\ \ref{f:gr2}). 
The maximum wavelength is obviously equal to $\Lambda_0=2\,\bar r$. \label{f:gr3}}
\end{figure}

{\bf Kolmogorov's constant {\mathversion{bold} $\alpha_1$}.}
This constant belongs to the wavenumber spectrum and deserves an idea about the 
outer limits and inner structure of an unstable, dissipative patch as sketched in 
Fig.\  \ref{f:gr3} for the begin of the cascade process evolving into a strucure 
like the one given in Fig.\ \ref{f:gr2}.

The most important message of Fig.\  \ref{f:gr3} is that the $longest$
or energy-containing wavelength of the dissipative patch equals 
$\Lambda_0=2\,\bar r$. The wavelength in a  dipole is $4\,\bar r$.
The dipole performs chaotic trajectories in a white-noise sense  and forms no patch or spectrum.
This difference between the two configurations is essential. We use our $\Lambda_0$ as
a lower integration limit for the spectral energy distribution. It is important to underline
that $\Lambda_0$ labels the upper wavelength  limit (the longest wavelike motions) 
in a dissipative patch. This limit is actually not influenced by the formation details of the spectrum. 

We integrate (\ref{kolmo1}) over the dissipatve patch
in the sense sketched in Figures \ref{f:gr2} and \ref{f:gr3} and get
\begin{equation}\label{kolmo2}
	{ K} = \alpha _1 \,\varepsilon^{2/3} \, \int _{k_{0}}^{\infty} k^{-5/3}\;dk 
	=\alpha _1\,\frac{3}{2} \,\left( \frac{\varepsilon}{k_{0}}\right)^{2/3},
\end{equation}
where $k_0=2\,\pi/\Lambda_0$ characterizes the lower end of the turbulence spectrum 
in the wavenumber space. We loosely assign the wavenumber range 
$k=0\dots k_0$ to the mean flow which may basically be resolved in numerical models.
The dissipation rate $\varepsilon$ in (\ref{kolmo2}) can be expressed as follows,
\begin{equation}\label{hilfs1}
	\varepsilon = {K}/\tau \,,
\end{equation}
with $\tau$ being the lifetime of a dissipative patch. 
Inserting (\ref{hilfs1}) in (\ref{kolmo2}) and rearranging gives the following:
\begin{equation}\label{kolmo13}
	\alpha_1 = \frac{2}{3}\,\left( 2\,\pi \right)^{2/3} {K}^{1/3} 
	\left(\frac{\tau}{2\,\bar r}\right)^{2/3}\,.
\end{equation}

In a local quasi-equilibrium sense for a dense vortex ensemble the marching dipoles 
can occupy only those places which are simultaneously also ``emptied'' through dipole
annihilation or dissipative patches by decay. This means that the life time of a dissipative 
patch, $\tau ={ K}/\varepsilon$, should equal the time of ``free flight'' of a dipole over a distance 
$2\,\bar r$:
\begin{equation}\label{quasistat}
	\tau= { K}/\varepsilon = 2\,\bar r/u\;.
\end{equation}
Here we used the scalar dipole velocity $u$,
\begin{equation}\label{dipvel}
	u=\omega\,r=\sqrt{2\,{ K}}.
\end{equation}
After some algebra we get the pre-factor of 
the  three-dimensional wavenumber spectrum as follows:
\begin{equation}\label{alpha1}
	\alpha _1\;=\;\frac{1 }{3}\,(4\,\pi )^{2/3}\;=\;1.802\,.
\end{equation}
The corresponding value of an ideal one-dimensional spectrum is one third 
of the above, i.e.\ 0.60.


\section{Discussion}

\noindent
{\bf Equations of turbulent motion.}
The results of our considerations can be summarized as follows:
\begin{eqnarray}
\label{K-111}
	\frac{\partial K}{\partial t} &+&\frac{\partial}{\partial \vec x}\left( \vec U\,K - \nu \frac{\partial K}{\partial \vec x} \right)
	=  \nu\left(S^2-\Omega^2\right),\\
\label{Om-111}
	\frac{\partial \Omega}{\partial t}&+& \frac{\partial}{\partial \vec x}\left( \vec U\,\Omega-\nu \frac{\partial\Omega}{\partial\vec x}\right)
	=\frac{1}{\pi} \left( \frac{S^2}{2} -\Omega^2\right),\\
\label{nu-55}
	\nu&=&\frac{K}{\pi\,\Omega}\,.
\end{eqnarray}
These equations are structurally  identical with the $k$-$\omega$ closure model discussed by 
\cite{wilcox06}. There are only slight differences in the pre-factors of the terms.

This theory applies exclusively to locally homogeneous, isotropic and 
moderately unsteady high-Reynolds number flows. Extreme non-stationarities and/or 
sharp spatial gradients like in shockwaves are possibly not covered.
As a rule, temporal changes of the mean flow should happen 
on time scales sufficiently long compared with $T=1/\Omega$ because otherwise
spectral universality (\ref{kolmo1}) has possibly not enough time 
to become well enough established.

Our equations reproduce the logarithmic law of the wall\footnote{The controversy 
of the logarithmic $versus$ a power-law boundary layer \cite[][]{barenblatt97} 
shall not be discussed here.} and predict the universal von-K\'{a}rm\'{a}n's constant 
as $\kappa = 1/\sqrt{2\,\pi} = 0.399\approx 0.4$ where the latter value counts 
as internationally accepted standard \cite[see also][]{hogstrom85}.
The value corresponds nicely to measurements under \textit{favorable} pressure gradients
\cite[][]{chauhanetal05}. Similar support comes from 
\cite{jimenezmoser07} on the basis of an extensive review. They state: 
\textit{The K{\'a}rm{\'a}n constant $\kappa \approx$ 0.4 is approximately universal.}

However, \cite{landaulifshitz_eng87} wrote on their p.\ 173 that
\textit{\dots $\kappa$ (is) a numerical constant, the \textit{von-K{\'a}rm{\'a}n constant}, whose value cannot be calculated theoretically 
and must be determined experimentally. It is found to be $\kappa = 0.4$.}

With respect to the measurability of $\kappa$ even the opposite might be true: 
if ``physics disappears'' when $Re\rightarrow\infty$ and only sort of ``inert geometry'' 
(the Euler equation) remains, $\kappa$ can possibly no longer be 
seen as a physical quantity as it represents pure geometry. 

But these interesting speculations shall not rise any doubt about the necessity
of superpipe facilities for very high-$Re$ number experiments 
like in Princeton \cite[][]{zagarolasmits98}, Oregon \cite[][]{donelly98}  or in the European 
CICLoPE  \cite[]{talamellietal09}.

Our equations describe the free decay of turbulence following $K \sim t^{-m}$ with $m=1$, in agreement with Dickey and Mellor's [1980] 
high-$Re$ laboratory experiments and with Oberlack's [2002] theoretical result for the Navier-Stokes equation. 
Today it is still not clear why some decay experiments lead to $m > 1$. Possibly it is a matter of 
initial conditions \cite[see][]{hurstvassilicos07}: at high $Re$ viscosity is comparatively
small so that its regularizing effect towards a finally more self-similar decay spectrum
will take more time than at lower $Re$. In some cases this time may exceed the lifetime of turbulence. 
This underlines the necessity of deeper experimental work.

{\bf Kolmogorov's constants.} 
The rounded numerical values $\alpha_1=1.8$ or 
$\alpha_1/3=0.6$ predicted by our theory for  Kolmogorov's universal constant 
are situated well within the error bars of many high-$Re$ observations, NSE and RG 
based analytical approximations, laboratory and DNS experiments. 

\begin{figure}[htb]
\centerline{\includegraphics[width=6cm,height=5cm,keepaspectratio]{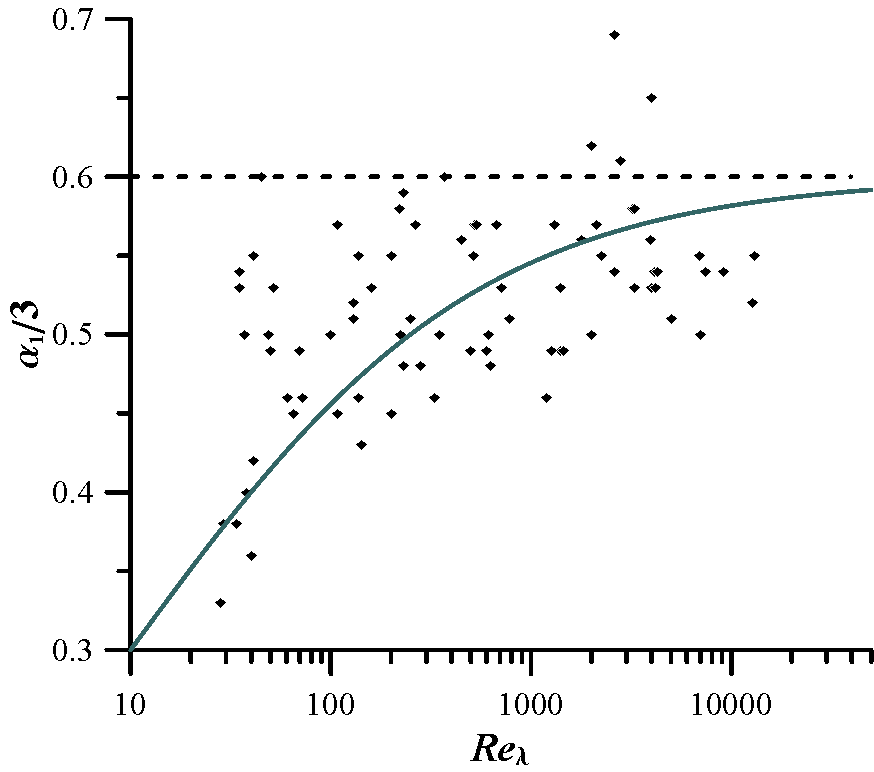}}
  \caption{Experimental and observational results  for $\alpha_1/3$
measured, collected from the literature, and analyzed by \cite{sreenivasan95}. 
The solid line follows our somewhat arbitrary approximation
$0.6\times {\sqrt{Re_{\lambda}}}/{\left( \sqrt{Re_*}+\sqrt{Re_{\lambda}}\right)}$ wherein
$\alpha_1/3=0.6$ is our theoretically derived  asymptotic value. Here we took $Re_{*}=10$.
\label{f:gr5}}
\end{figure}

Based on observations, \cite{tennekeslumley72} gave the value 
$\alpha _1 = 1.62$, but still with some uncertainty.
The study by \cite{sreenivasan95} (see Fig.\ \ref{f:gr5}) is possibly the most comprehensive 
review of experimental and observational values for the number $\alpha _1/3 $ until now.
Later \cite{yeungzhou97} reported a value of $\alpha_1 = 1.62$ based on high-resolution 
DNS studies with up to 512$^3$ grid points. A most recent study by 
\cite{donzissreenivasan10} on a DNS grid of $4096^3$ gave $\alpha_1\approx 1.58$. 

\noindent
Much higher Reynolds numbers  than in DNS could be achieved in oceanic measurements 
of Lagrangian frequency spectra by  \cite{liendasaro02}. These authors stated for the 
prefactor $\beta_1$ in the Lagrangian frequency spectrum that 
\textit{  \dots since the present uncertainty
is comparable to that between high quality estimates of the 
Eulerian one-dimensional longitudinal Kolmogorov constant measured by many dozen investigators
over the last 50 years, large improvements in the accuracy of the estimate of 
$\beta _1$ seem unlikely.}
For completeness reasons we mention other theoretical efforts to calculate the universal constants. 
They are technically extremely complex and neither unique nor part of an integrated descriptive concept 
for turbulence so that they all carry more singular characters. Beginning with an initiating work by \cite{forsteretal77}, systematic analytical approximations 
using RG methods and related techniques for NSE gave rise to some estimates. 
E.g.\ \cite{yakhotorszag86a,yakhotorszag86b} found  $\alpha_1\approx 1.62$ 
whereas \cite{mccombwatt92} derived $\alpha_1=1.60\pm0.01$ and \cite{parkdeem03}
obtained $\alpha_1 = 1.68$. 

{\bf Coda.}
Saffman  [1977] \cite[loc. cit.][p.\ 107]{davidson04} feared that 
\textit{\dots in searching for a theory of turbulence, 
perhaps we are looking for a chimera.}  

This has recently been enforced by \cite{hunt2011}: 
\textit{\dots But there are good reasons why the  answer to the  
big  question that  Landau and Batchelor raised about whether there  
is a general theory of  turbulence is probably {`no'}.}  

Due to our new theory which draws some profit namely from 
the two-fluid ideas of Landau we are more optimistic. But
skeptical thoughts remain in view  of the extraordinary low 
measurement accuracy of the universal constants of turbulence 
compared with the extremely high precisision of the 
fundamental constants of physics like e.g.\ 
the vacuum speed of light, or the mass of the 
proton \cite[]{fritzsch09}. 
The challenge is possibly the dynamic character of turbulence which is 
in the best case characterized as a stable steady state rather than
a static property like the electron's electric charge. 

\begin{acknowledgments}
This work profited from a fruitful cooperation with Hartmut Peters at 
Earth and Space Research in Seattle, WA, as part of 
the Department of the Navy Grant N62909-10-1-7050 issued by the 
Office of Naval Research Global. The United States Government has 
a royalty-free license throughout the world in all copyrightable material contained herein. 

The author thanks Eric D'Asaro, Bruno Eckhardt, Michael Eckert, Philippe Fraunier, 
Eckhard Kleine, Rupert Klein, Jim Riley, J\"urgen S\"undermann,  Ed Spiegel,
Oleg F. Vasiliev, and Michael Wilczek. He thanks the  organizers of 
\textit{Turbulent Mixing and Beyond 2011} at the Abdus Salam ICTP in 
Trieste/Italy for creating there a fair, open and creative atmosphere. 
\end{acknowledgments}

\bibliography{baumert-final-bib1.bib,baumert-final-bib2.bib}
\end{document}